\documentclass{article}

\usepackage[T1]{fontenc} 
\usepackage[margin=1.1in]{geometry}
\usepackage{graphicx}
\usepackage[minbibnames=1, maxbibnames=99]{biblatex}
\usepackage{amsmath,amsfonts,amssymb}
\usepackage{hyperref}
\hypersetup{
    colorlinks=true,
    linkcolor=blue,
    filecolor=black,      
    urlcolor=blue,
    citecolor=black,
}

\usepackage{float}
\usepackage{booktabs}
\usepackage[table,xcdraw]{xcolor}
\usepackage{eurosym}
\usepackage{multirow}
\usepackage[none]{hyphenat}
\usepackage[misc]{ifsym}

\title{\Large\textbf{Artificial Intelligence-Based Classification of Spitz Tumors}}
\author{Ruben T. Lucassen\,$^{1,2,}$\textsuperscript{\Letter}, Marjanna Romers\,$^{1}$, Chiel F. Ebbelaar\,$^{1,3}$, Aia N. Najem\,$^{1,4}$,\\ Donal P. Hayes\,$^{5}$, Antien L. Mooyaart\,$^{6}$, Sara Roshani\,$^{7}$, Liliane C. D. Wynaendts\,$^{8}$\\ Nikolas Stathonikos\,$^{1}$, Gerben E. Breimer\,$^{1}$, Anne M. L. Jansen\,$^{1}$,\\ Mitko Veta\,$^{2,*}$, Willeke A. M. Blokx\,$^{1,*}$}
\date{}

\begin{document}

\newcommand{\red}[1]{\textcolor{red}{#1}}
\newcommand{\blue}[1]{\textcolor{blue}{#1}}
\newcommand{\etal}{~\textit{et al.}}
\newcommand\nomarkerfootnote[1]{
  \begingroup
  \renewcommand\thefootnote{}\footnote{#1}%
  \addtocounter{footnote}{-1}%
  \endgroup
}

\maketitle

\noindent$^{1}$~Department of Pathology, University Medical Center Utrecht, Utrecht, the Netherlands\\
$^{2}$~Department of Biomedical Engineering, Eindhoven University of Technology, Eindhoven, the Netherlands\\
$^{3}$~Department of Dermatology, Leiden University Medical Center, Leiden, the Netherlands\\
$^{4}$~Department of Pathology, Radboud University Medical Center, Nijmegen, the Netherlands\\
$^{5}$~Department of Pathology, Meander Medical Center, Amersfoort, the Netherlands\\
$^{6}$~Department of Pathology, Erasmus Medical Center, Rotterdam, the Netherlands\\
$^{7}$~Department of Pathology, Amsterdam University Medical Center, Amsterdam, the Netherlands\\
$^{8}$~Department of Pathology, St. Antonius Hospital, Nieuwegein, the Netherlands\\

\noindent\textsuperscript{\Letter}~Corresponding author\\
Ruben T. Lucassen, MSc\\
Department of Pathology\\
University Medical Center Utrecht\\
Heidelberglaan 100, 3584 CX, Utrecht, the Netherlands\\
E-mail Address: r.t.lucassen@umcutrecht.nl\\
\\
\noindent\textsuperscript{*}~These authors contributed equally: Mitko Veta and Willeke A. M. Blokx

\newpage

\section*{Abstract}
Spitz tumors are diagnostically challenging due to overlap in atypical histological features with conventional melanomas. We investigated to what extent AI models, using histological and/or clinical features, can: (1) distinguish Spitz tumors from conventional melanomas; (2) predict the underlying genetic aberration of Spitz tumors; and (3) predict the diagnostic category of Spitz tumors. 
The AI models were developed and validated using a retrospective cohort from the University Medical Center Utrecht, the Netherlands. The dataset consisted of 393 Spitz tumors and 379 conventional melanomas. Predictive performance was measured using the area under the receiver operating characteristic curve (AUROC) and the accuracy. The performance of the AI models was compared with that of four experienced pathologists in a reader study. Moreover, a simulation experiment was conducted to investigate the impact of implementing AI-based recommendations for ancillary diagnostic testing on the workflow of the pathology department.
The best AI model based on UNI features reached an AUROC of 0.95 (95\% CI, 0.92-0.98) and an accuracy of 0.86 (95\% CI, 0.81-0.91) in differentiating Spitz tumors from conventional melanomas. The genetic aberration was predicted with an accuracy of 0.55 (95\% CI, 0.46-0.64) compared to 0.25 for randomly guessing. The diagnostic category was predicted with an accuracy of 0.51 (95\% CI, 0.40-0.60), where random chance-level accuracy equaled 0.33. On all three tasks, the AI models performed better than the four pathologists, although differences were not statistically significant for most individual comparisons. Based on the simulation experiment, implementing AI-based recommendations for ancillary diagnostic testing could reduce material costs, turnaround times, and examinations.
In conclusion, the AI models achieved a strong predictive performance in distinguishing between Spitz tumors and conventional melanomas. On the more challenging tasks of predicting the genetic aberration and the diagnostic category of Spitz tumors, the AI models performed better than random chance.

\newpage

\section*{Introduction}
Cutaneous melanocytic lesions are categorized into many subtypes, each with distinct biological behavior~\cite{who2023}. One of these subtypes, known as Spitz tumors, mostly develops at a young age and is histologically characterized by the presence of large epithelioid and/or spindled melanocytes with variable cytonuclear atypia~\cite{harms2015atypical}. Similar atypia is also frequently seen in conventional melanomas, making it challenging at times to differentiate the two based on histopathological assessment alone, as evidenced by only a moderate inter-observer agreement between expert dermatopathologists~\cite{benton2021impact}. Whereas conventional melanomas are by definition malignant, the majority of Spitz tumors display benign biological behavior. For this reason, there is a high risk of both under- and overtreatment in case of misdiagnosis~\cite{harms2015atypical}. Immunohistochemical (IHC) staining and molecular analyses can often alleviate the diagnostic challenge by identifying a defining genetic aberration (i.e., a \textit{BRAF} or \textit{NRAS} mutation in conventional melanomas and an \textit{HRAS} mutation or kinase fusion in Spitz tumors)~\cite{bastian2014molecular}, but are more expensive and time-consuming to perform.

Recent advances in artificial intelligence (AI) show promising results for a range of diagnostic and prognostic applications in pathology~\cite{song2023artificial,shmatko2022artificial}. Several studies have explored the use of AI models for classification of Spitz tumors using learned or human-interpreted features from whole slide images (WSIs), but were mainly limited by small datasets and a lack of genetic confirmation of the defining driver aberration for all lesions included~\cite{hart2019classification,snyder2022histologic,mosquera2024histological}. In this study, we investigate the accuracy with which an AI model, using histological and/or clinical features, can perform three prediction tasks: (1) distinguishing Spitz tumors from conventional melanomas; (2) predicting the underlying genetic aberration of Spitz tumors (i.e., a fusion in \textit{ALK}, \textit{ROS1}, \textit{NTRK}, or all other Spitz-related aberrations); and (3) predicting the diagnostic category of Spitz tumors (i.e., benign, intermediate, or malignant). We conduct a reader study to compare the performance of the AI models with that of four experienced pathologists. Moreover, to study how implementing AI-based recommendations for ancillary diagnostic testing could affect the workflow of the pathology department, we conduct a simulation experiment. While perfect predictive performance for all of these tasks is unlikely based on histological and clinical features alone, even an AI model with reasonable performance can potentially be valuable, for example as a decision-support tool for guiding pathologists in the selection of ancillary diagnostic tests to reach the correct diagnosis more efficiently.

\section*{Methods}
\subsection*{Study design}
This retrospective cohort study was performed using archival data from the pathology department of the University Medical Center (UMC) Utrecht, the Netherlands. All genetically confirmed conventional melanomas and Spitz tumors accessioned between January 1, 2013, and August 31, 2023, were included. The study does not fall within the scope of the Dutch Medical Research Involving Human Subjects Act (WMO) and therefore does not require approval from an accredited medical ethics committee in the Netherlands. Nevertheless, an independent quality assessment (25U-0162) was conducted at the UMC Utrecht to ensure compliance with relevant laws and regulations, including those related to the informed consent procedure, data management, privacy, and legal considerations. All data were pseudonymized. Data from patients who opted out of the use of their data for research purposes were excluded.

\begin{table}[]
\centering
\caption{Patient and lesion characteristics for Spitz tumors and conventional melanomas.}
\begin{tabular}{@{}llrrrrl@{}}
\toprule\toprule
 & Characteristics          & \multicolumn{2}{c}{\begin{tabular}[c]{@{}c@{}}Spitz\\ Tumors\end{tabular}}   & \multicolumn{2}{c}{\begin{tabular}[c]{@{}c@{}}Conventional\\ Melanomas\end{tabular}} &  \\ 
 &                          & \multicolumn{2}{c}{(N = 393)} & \multicolumn{2}{c}{(N = 379)}                                         &  \\\midrule
 & \textbf{Age}             &      & \multicolumn{1}{l}{} &                                  & \multicolumn{1}{l}{}                             &  \\
 & \hspace{4mm}Median (IQR)             & \multicolumn{2}{l}{27 (16)} & \multicolumn{2}{l}{48 (28)}                                             &  \\
 & \hspace{4mm}Min-Max                  & \multicolumn{2}{l}{1-73}    & \multicolumn{2}{l}{3-85}                                                &  \\
 & \textbf{Sex (\%)}             &      & \multicolumn{1}{l}{} &                                  & \multicolumn{1}{l}{}                             &  \\
 & \hspace{4mm}Male                     & 118  & (30.0)               & 158                              & (41.7)                               &  \\
 & \hspace{4mm}Female                   & 275  & (70.0)               & 221                              & (58.3)                               &  \\
 & \textbf{Location (\%)}        &      & \multicolumn{1}{l}{} &                                  & \multicolumn{1}{l}{}                             &  \\
 & \hspace{4mm}Head and neck            & 32   & (8.1)                & 56                               & (14.8)                               &  \\
 & \hspace{4mm}Trunk                    & 73   & (18.6)               & 154                              & (40.6)                               &  \\
 & \hspace{4mm}Upper extremities        & 66   & (16.8)               & 55                               & (14.5)                               &  \\
 & \hspace{4mm}Lower extremities        & 197  & (50.1)               & 94                               & (24.8)                               &  \\
 & \hspace{4mm}Hands and feet           & 23   & (5.9)                & 13                               & (3.4)                                &  \\
 & \hspace{4mm}Unknown                  & 2    & (0.5)                & 7                                & (1.8)                                &  \\
 & \textbf{Diagnostic category (\%)}         &      & \multicolumn{1}{l}{} &                                  & \multicolumn{1}{l}{}                             &  \\
 & \hspace{4mm}Benign                   & 209  & (53.2)               & \multicolumn{2}{c}{-}                                                   &  \\
 & \hspace{4mm}Benign / intermediate    & 37   & (9.4)                & \multicolumn{2}{c}{-}                                                   &  \\
 & \hspace{4mm}Intermediate             & 95   & (24.2)               & \multicolumn{2}{c}{-}                                                   &  \\
 & \hspace{4mm}Intermediate / malignant & 17   & (4.3)                & \multicolumn{2}{c}{-}                                                   &  \\
 & \hspace{4mm}Malignant                & 35   & (8.9)                & 379                              & (100.0)                              &  \\
 & \textbf{Genetic aberration (\%)}       &      & \multicolumn{1}{l}{} &                                  & \multicolumn{1}{l}{}                             &  \\
 & \hspace{4mm}Mutations                &      & \multicolumn{1}{l}{} &                                  & \multicolumn{1}{l}{}                             &  \\
 & \hspace{8mm}\textit{BRAF}            & \multicolumn{2}{c}{-}       & 263                              & (69.4)                                           &  \\
 & \hspace{8mm}\textit{NRAS}            & \multicolumn{2}{c}{-}       & 112                              & (29.6)                                           &  \\
 & \hspace{8mm}\textit{BRAF}\,\&\,\textit{NRAS}  & \multicolumn{2}{c}{-}       & 4                                & (1.1)                                            &  \\
 & \hspace{8mm}\textit{HRAS}            & 34   & (8.7)                & \multicolumn{2}{c}{-}                                                               &  \\
 & \hspace{8mm}\textit{ROS1}            & 1    & (0.3)                & \multicolumn{2}{c}{-}                                                               &  \\
 & \hspace{4mm}Fusions                  &      & \multicolumn{1}{l}{} &                                  & \multicolumn{1}{l}{}                             &  \\
 & \hspace{8mm}\textit{ROS1}            & 106  & (27.0)               & \multicolumn{2}{c}{-}                                                               &  \\
 & \hspace{8mm}\textit{NTRK}            & 111  & (28.2)               & \multicolumn{2}{c}{-}                                                               &  \\
 & \hspace{12mm}\textit{NTRK1}          & 17   & (4.3)                & \multicolumn{2}{c}{-}                                                              &  \\
 & \hspace{12mm}\textit{NTRK2}          & 31   & (7.9)                & \multicolumn{2}{c}{-}                                                              &  \\
 & \hspace{12mm}\textit{NTRK3}          & 27   & (6.9)                & \multicolumn{2}{c}{-}                                                              &  \\
 & \hspace{12mm}Unknown                 & 36   & (9.2)                & \multicolumn{2}{c}{-}                                                              &  \\
 & \hspace{8mm}\textit{ALK}             & 59   & (15.0)               & \multicolumn{2}{c}{-}                                                               &  \\
 & \hspace{8mm}\textit{MAP3K8}          & 41   & (10.4)                & \multicolumn{2}{c}{-}                                                               &  \\
 & \hspace{8mm}\textit{BRAF}            & 18   & (4.6)                & \multicolumn{2}{c}{-}                                                               &  \\
 & \hspace{8mm}\textit{RET}             & 18   & (4.6)                & \multicolumn{2}{c}{-}                                                               &  \\
 & \hspace{8mm}\textit{MET}             & 4    & (1.0)                & \multicolumn{2}{c}{-}                                                               &  \\
 & \hspace{8mm}\textit{RASGFR1}         & 1    & (0.3)                & \multicolumn{2}{c}{-}                                                               &  \\ 
 & \textbf{WSI availability (\%)}            &      & \multicolumn{1}{l}{} &                                  & \multicolumn{1}{l}{}                             &  \\
 & \hspace{4mm}Internal and consultation & 264 & (67.2)            & 220                              & (58.0)                                          &  \\
 & \hspace{4mm}Internal only            & 102  & (26.0)            & 117                              & (30.9)                                          &  \\
 & \hspace{4mm}Consultation only        & 27   & (6.9)             & 42                               & (11.1)                                          &  \\ \bottomrule\bottomrule
\end{tabular}
\label{tab:characteristics}
\end{table}

\subsection*{Dataset curation}
A total of 772 primary cutaneous melanocytic lesions were included in the dataset, comprising 379 conventional melanomas and 393 Spitz tumors (including nevi, melanocytomas, and melanomas). The lineage of the lesions (i.e., Spitz or conventional melanoma) was confirmed using IHC staining, fluorescence in situ hybridization (FISH), next generation sequencing (NGS), and/or targeted RNA sequencing. Lesions without a confirmed lineage were excluded. The diagnostic category of the lesions was determined based on histological features, IHC stains (e.g., PRAME and p16 expression), and genetic aberrations (e.g., the number of segmental copy number variations determined using SNP array analysis, absence or presence of secondary pathogenic mutations in for instance the \textit{TERT} promotor or in the \textit{TP53} or \textit{CDKN2A} gene). A pre-existing nevus was observed in 21.1\% of the conventional melanomas. The majority of the included lesions (80.4\%) concerned referral cases for consultation. Hence, for most lesions there were WSIs available of slides prepared at the referring center and internal slides prepared at the pathology department of the UMC Utrecht, with a different hematoxylin and eosin (H\&E) appearance due to variation in preparation and staining protocols. Characteristics of the lesions in the dataset are summarized in Table~\ref{tab:characteristics}. 

The tissue specimens consisted of shave and punch biopsies, excisions, and re-excisions. If multiple specimens of the same lesion were available, for example in case of an initial biopsy followed by a re-excision with lesion tissue remaining, the WSIs were grouped at the lesion level. All WSIs of unique, H\&E-stained slides with lesion tissue present were included per lesion. Image acquisition was performed using a ScanScope XT scanner (Aperio, Vista, CA, USA) at 20$\times$ magnification with a resolution of 0.50 \textmu m per pixel (slides scanned before 2016), a NanoZoomer 2.0-XR scanner (Hamamatsu photonics, Hamamatsu, Shizuoka, Japan) at 40$\times$ magnification with a resolution of 0.23 \textmu m per pixel (slides scanned starting from 2016 until May 2022), and a NanoZoomer S360 scanner (Hamamatsu photonics, Hamamatsu, Shizuoka, Japan) at 40$\times$ magnification with a resolution of 0.23 \textmu m per pixel (slides scanned after May 2022).

The dataset was randomly split at the patient level into a model development set (75\%) and test set for evaluation (25\%). The development set was further subdivided into five folds for cross-validation. To investigate the model performance subject to variation in H\&E staining, only lesions with both WSIs of internal and consultation slides were sampled for inclusion in the evaluation set, this while maintaining a prevalence of lesion (sub)types comparable to the development set.

\subsection*{Feature representation}
\label{sec:feature_representation}
Tissue cross-sections and pen markings were segmented in each WSI at 1.25$\times$ magnification using SlideSegmenter~\cite{lucassen2024tissue}. The resulting tissue segmentation map was used to guide the slide tessellation. Non-overlapping image tiles were extracted from the tissue regions of the WSIs at 20$\times$ magnification. Tiles mostly showing the uninformative background of the slide (i.e., for less than 5\% covered by tissue) and tiles showing pen markings were excluded. The remaining image tiles were converted into feature vectors, capturing the visual information in a compressed form to reduce the computational demands for analysis. Feature vectors were extracted for all tissue tiles using three different feature encoders: (1) First stage of HIPT~\cite{chen2022scaling} producing 192-dimensional feature vectors for tiles of 256$\times$256 pixels; (2) Second stage of HIPT producing 384-dimensional feature vectors for tiles of 4,096$\times$4,096 pixels; and (3) UNI~\cite{chen2024uni} producing 1024-dimensional feature vectors for tiles of 224$\times$224 pixels.

\subsection*{Model training}
AI models were trained for each of the three classification tasks using the three sets of extracted feature vectors. Across all combinations of the task and feature vector set, model training was repeated five times, each using a different fold for validation and the remaining four folds for training. Since the number of extracted feature vectors varies per case, only feature vectors from a single case were used per iteration (i.e., a batch size of one). The Vision Transformer (ViT)~\cite{dosovitskiy2020vit} (depth~=~2, heads~=~4, MLP-ratio~=~4, embedding dimension~=~192) was used as model architecture. The models were trained by minimizing the cross-entropy loss for 32,000 iterations starting from randomly initialized parameters using the AdamW~\cite{loshchilov2019decoupled} optimization algorithm ($\beta_1$~=~0.9, $\beta_2$~=~0.999). To counteract the class imbalance in the diagnostic category prediction task, which was not as severe for the other two tasks, the models were optimized with balancing class weights for this task. Gradients were accumulated over every 32 iterations. The learning rate was 5\,$\cdot$\,10$^{-5}$ at the start and halved after every 6,400 iterations. The network parameters that resulted in the smallest loss on the validation fold were saved, which was evaluated after every 320 iterations. The models were trained with attention dropout (p~=~0.5). In addition, feature vectors were randomly excluded during training as another form of dropout (p~=~0.5). If the total number of features for a case exceeded the maximum of 25,000 feature vectors, a subset equal in size to the maximum was randomly selected. Hyperparameters were tuned based on the average performance on the five validation folds. The predicted probability threshold for the binary classification task was optimized based on the performance on the validation set for each model. For the classification tasks with more than two classes, the class with the largest predicted probability was considered to be the predicted class. The model as well as the training and evaluation procedure were implemented in the Pytorch~\cite{paszke2019pytorch} framework. The code and trained model parameters are made publicly available\footnote{\url{https://github.com/RTLucassen/spitz_classification}}.

\subsection*{Experimental setup}
For the Spitz classification tasks, we compared three approaches: (1) logistic regression using clinical features only (i.e., age, sex, and anatomical location); (2) ViTs using image features only (based on the first and second stage of HIPT as well as UNI); (3) logistic regression using the clinical features in combination with the image-based feature vector extracted before the final layer of the ViTs. Because some Spitz tumors harbor rare genetic aberrations, not enough cases were available to form a separate class for development and evaluation of the AI models, which is why the cases were grouped for classification into an ALK, ROS1, NTRK, and other class. This aligns well with the fact that the IHC stains for ALK, ROS1, and NTRK are also the most widely available and commonly used for Spitzoid lesions. Similarly, Spitz tumors with a differential diagnosis of benign/intermediate or intermediate/malignant as diagnostic category were grouped with the more severe category for classification. The predicted probability for individual cases with more than the maximum of 25,000 feature vectors was considered to be the average of the predicted probabilities based on 10 randomly selected subsets of the maximum size. Probabilities predicted by the five model instances developed in the cross-validation were averaged to obtain model ensemble predictions. Model performance was measured in terms of the area under the receiver operating characteristic curve (AUROC) and accuracy on both the internal and consultation test set. The AUROC for multi-class classification tasks was computed per class using a one-versus-rest approach. Stratified bootstrapping (R = 10,000 samples) was used to calculate 95\% confidence intervals (CIs) using the percentile method. A binomial test was used to statistically compare the accuracy of the best AI model to the expected accuracy when randomly guessing. A \textit{P} value below 0.05 was considered statistically significant.

\subsection*{Reader study}
We conducted a reader study to compare the performance of the best AI models with that of pathologists' assessment on the three classification tasks. We recruited two pathologists from different academic centers and two pathologists from non-academic centers, all of whom had five or more years of experience in dermatopathology. A stratified subset of 100 cases was randomly selected from the internal test set. The reader study was performed using the SlideScore platform\footnote{\url{www.slidescore.com}} where the pathologists were provided with the most representative WSI per case and the corresponding clinical information. The participating pathologists were blinded from any additional diagnostic information (e.g., IHC-stained slides or findings from molecular analyses). Only if a case was classified as a Spitz tumor by the pathologist, the questions related to the genetic aberration and diagnostic category appeared and could be answered in the user interface. The order in which the cases were presented was randomized. For a fair comparison, we also evaluated the best AI model on the subset of selected cases using only the most representative WSI, different from before where all WSIs with tumor tissue present were provided. McNemar's exact test~\cite{mcnemar1947note} was used for statistical comparison between the accuracy of each pathologist and the best AI models on the three tasks. The Bonferroni-correction was applied to adjust the \textit{P} values for multiplicity (4 comparisons). Since the genetic aberrations and diagnostic categories were only predicted by the pathologists when a lesion was first identified as a Spitz tumor, the statistical comparison of the accuracy for these two tasks was limited to subset of true Spitz tumors with pathologists' predictions available, which differed for each pathologist. The corresponding AI model predictions were selected for each subset to allow for paired comparisons.

\begin{figure}[h!]
    \centering
    \includegraphics[width=0.62\textwidth]{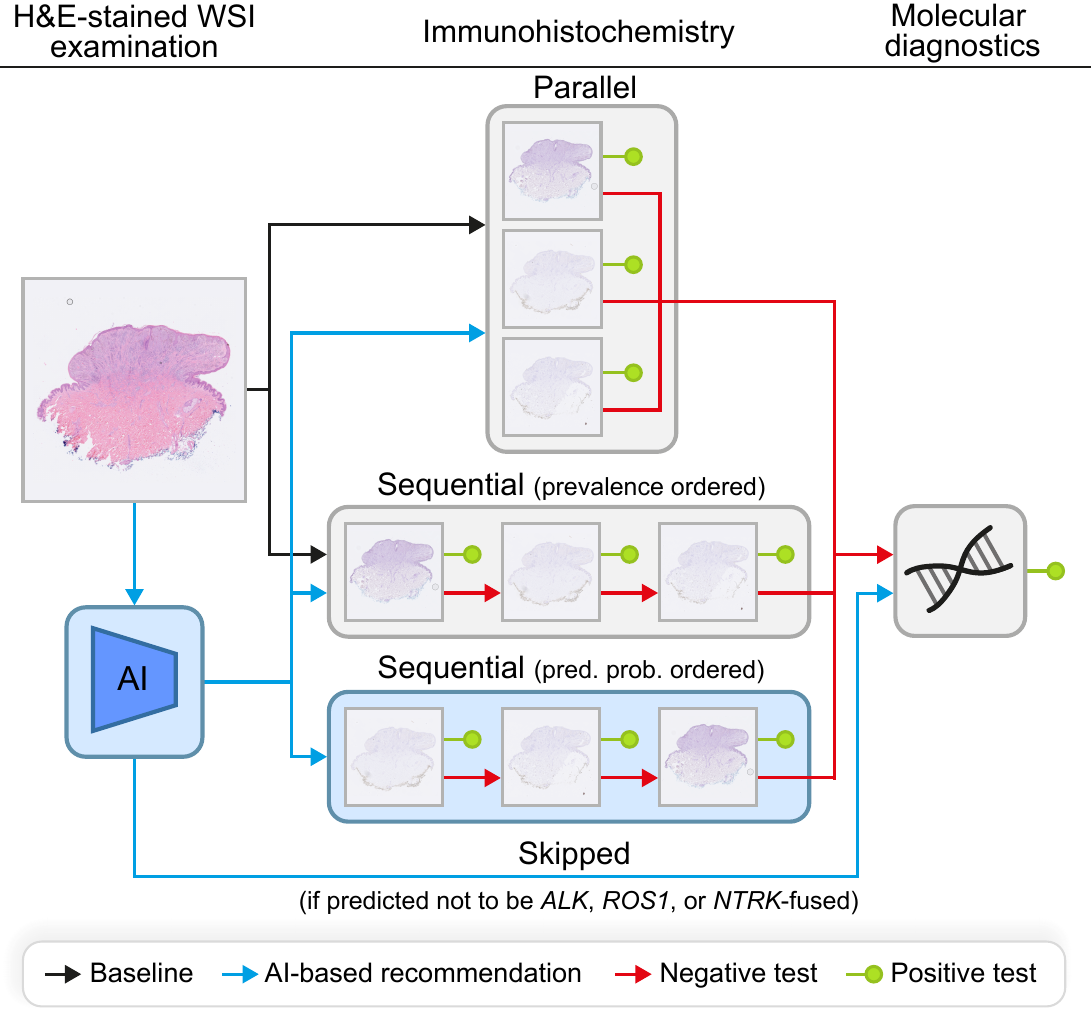}
    \caption{Flowchart of the baseline and AI-incorporated workflow variants for the simulation experiment. In the baseline workflow, IHC staining is either performed in parallel or in sequence ordered from high to low prevalence. In the workflow with AI-based recommendations based on the predicted probabilities for the genetic aberrations, IHC staining is either skipped, or performed in parallel, in sequence ordered from high to low prevalence, or in sequence ordered from high to low predicted probability (abbreviated as pred. prob.).}
    \label{fig:workflow}
\end{figure}

\subsection*{Simulation experiment}
A simulation experiment was conducted to investigate how implementing AI-based recommendations of ancillary diagnostic tests based on the predicted genetic background of Spitz tumors could affect the workflow of the pathology department. A flowchart of the simulated workflow variants is shown in Fig.~\ref{fig:workflow}. The typical workflow at the pathology department of the UMC Utrecht starts with performing the Spitz (i.e., ALK, ROS1, and NTRK) IHC stains, which are followed by molecular diagnostics if necessary. In the simulation, as soon as a positive IHC stain is identified, potentially remaining IHC stains and molecular diagnostics are not performed anymore. Two baseline variants were defined, with IHC stains performed either in parallel or sequentially, ordered from high to low prevalence of the corresponding genetic aberration. The baselines were also expanded by incorporating AI-based recommendations. If the AI model classifies a lesion to be part of the class with other Spitz tumors (i.e., not \textit{ALK}, \textit{ROS1}, and \textit{NTRK}-fused) with a predicted probability that exceeds the threshold $T$, IHC staining is skipped and molecular analysis is performed directly. In addition, the order of the sequential IHC stains can alternatively be based on the probabilities predicted by the AI model instead of the prevalence. To put the results of the best AI model for genetic aberration prediction into perspective, the simulation was also repeated with a hypothetical perfect AI-based recommendation system.

All simulated workflow variants were repeated for 10,000 iterations. Per iteration, 100 Spitz cases were randomly sampled with replacement from the test set, which approximately reflects the number of genetically confirmed Spitz cases diagnosed annually in the pathology department of the UMC Utrecht. The Spitz IHC stains were assumed to cost \euro100 each~\cite{ebbelaar2024comparative} and to require 1 day of processing time. Molecular diagnostics was assumed to cost \euro1000~\cite{hindi2020feasibility,wolff2022cost} and to require 10 days of processing time. The assumed costs and turnaround times were based on our experience at UMC Utrecht and values reported in the literature, but may vary between centers. False negative or ambiguous IHC stains are not uncommon in practice and were incorporated in the simulation. Based on the proportions in the complete dataset, the probabilities of an ALK, ROS1, and NTRK IHC stain being false negative or too ambiguous for definitive diagnosis in the simulation were 0.055, 0.448, 0.255, respectively. Empirically, we found $T$~=~0.5 to be a suitable threshold for the AI model we developed. The simulation results include the mean and 95\% CI of the material cost accumulated over 100 cases, the average turnaround time per case, and the average number of examinations by a pathologist because of new diagnostic information per case (e.g., an initial examination of H\&E-stained slides, followed by re-examination after the IHC-stained slides have been prepared, followed by another re-examination after the results of molecular analyses are available, equals three examinations in total).

\section*{Results}
\subsection*{Spitz Tumor versus Conventional Melanoma Prediction}
The test set results of the prediction models for distinguishing Spitz tumors from conventional melanomas are shown in Table~\ref{tab:spitz_vs_conv_mel}. The logistic regression model based only on clinical features achieved an AUROC of 0.80 (95\% CI, 0.74-0.86) and an accuracy of 0.74 (95\% CI, 0.66-0.79). In comparison, all AI models based only on image-extracted features performed better than the clinical model. Using the second-stage features of HIPT resulted in slightly higher performance scores than using the features after the first stage of HIPT. The best performance was obtained by the AI model based on the features extracted using UNI with an AUROC of 0.95 (95\% CI, 0.92-0.98) and an accuracy of 0.86 (95\% CI, 0.81-0.91) using the internal WSIs, which was statistically significantly different (\textit{P} < 0.001) from the expected accuracy of 0.50 for random predictions. Out of the seven Spitz tumors incorrectly classified by the model as conventional melanomas, three were benign Spitz nevi, three were Spitz melanocytomas, and one was a Spitz melanoma. Overall, the performance was slightly better when evaluated on the internal WSIs than on the consultation WSIs. Combining the best image-extracted features with the clinical features resulted in comparable performance.

\begin{figure}[h!]
    \centering
    \includegraphics[width=0.62\textwidth]{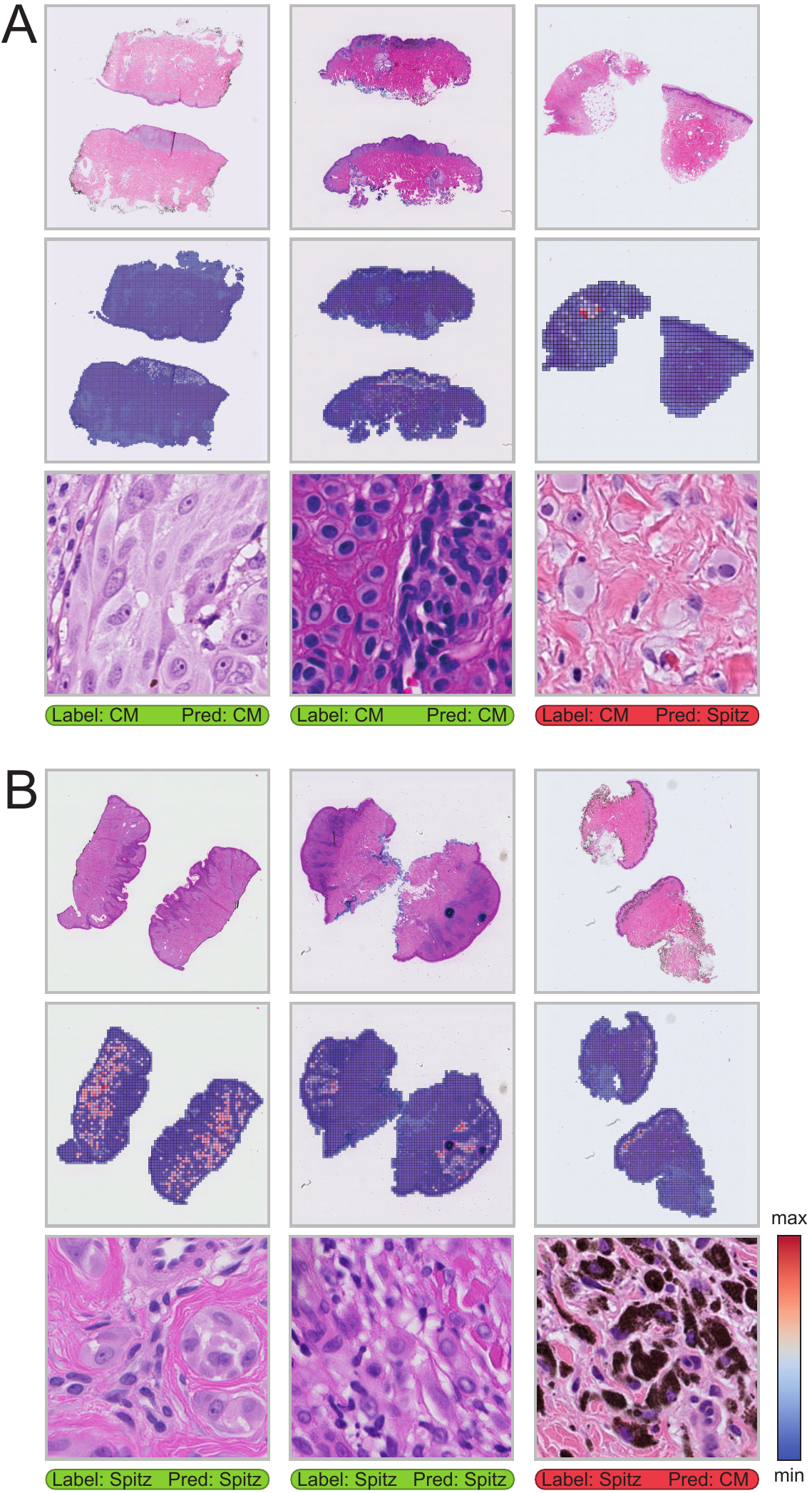}
    \caption{Example cases from the test set. Per case from top to bottom: tissue cross-sections from the most representative whole slide image for that case, the tiles extracted from the cross-sections (excluding pen markings) colored based on the attention weights assigned by the AI model, the tile with the largest attention weight at a higher magnification, and the classification result. Classification decisions were obtained using the best threshold based on the validation fold. (A) Predictions for conventional melanoma (CM) cases. (B) Predictions for Spitz tumor cases.}
    \label{fig:attention_maps}
\end{figure}

Several example cases with corresponding attention maps and classification results for one of the five UNI features-based models in the ensemble are shown in Fig.~\ref{fig:attention_maps}. The attention maps highlight the importance of each tile for the case-level prediction by way of the model-assigned weight. Tiles that were assigned the highest attention weight consistently showed the melanocytic lesion, primarily the dermal component, for both correct and incorrect predictions. Moreover, in conventional melanoma cases with a pre-existing nevus, the nevus tiles often received high attention weights (see center column of Fig.~\ref{fig:attention_maps}A).

\begin{table}[h!]
\centering
\caption{Results for the Spitz tumor versus conventional melanoma prediction on the test set.}
\resizebox{\columnwidth}{!}{%
\begin{tabular}{llcccc}
\toprule\toprule
Features          & Feature extractor & \multicolumn{2}{c}{Internal WSIs} & \multicolumn{2}{c}{Consultation WSIs}                                    \\ \cmidrule(lr){3-4}\cmidrule(lr){5-6}
                  &                   & AUROC (95\% CI) & Acc. (95\% CI) & AUROC (95\% CI) & Acc. (95\% CI) \\ \hline
Clinical only     & -                 & 0.80~~{\scriptsize(0.74-0.86)} & 0.74~~{\scriptsize(0.68-0.80)} & 0.80~~{\scriptsize(0.74-0.86)} & 0.74~~{\scriptsize(0.68-0.80)} \\
Image only        & HIPT (stage 1)    & 0.84~~{\scriptsize(0.78-0.90)} & 0.77~~{\scriptsize(0.71-0.83)} & 0.82~~{\scriptsize(0.76-0.87)} & 0.75~~{\scriptsize(0.69-0.80)} \\
                  & HIPT (stage 2)    & 0.87~~{\scriptsize(0.82-0.92)} & 0.79~~{\scriptsize(0.73-0.84)} & 0.85~~{\scriptsize(0.79-0.90)} & 0.74~~{\scriptsize(0.68-0.80)} \\
                  & UNI               & 0.95~~{\scriptsize(0.92-0.98)} & 0.86~~{\scriptsize(0.81-0.91)} & 0.93~~{\scriptsize(0.90-0.96)} & 0.85~~{\scriptsize(0.80-0.90)} \\
Clinical \& Image & UNI               & 0.95~~{\scriptsize(0.92-0.98)} & 0.86~~{\scriptsize(0.81-0.91)} & 0.94~~{\scriptsize(0.91-0.97)} & 0.85~~{\scriptsize(0.80-0.90)} \\
\bottomrule\bottomrule
\multicolumn{6}{l}{Acc. = Accuracy}\\
\end{tabular}
}
\label{tab:spitz_vs_conv_mel}
\end{table}

\newpage
\newpage

\subsection*{Spitz Genetic Aberration Prediction}
The best results for the prediction of the genetic aberrations in Spitz tumors were achieved using features extracted with UNI and are shown in Table~\ref{tab:genetic_aberration}. The AI model reached an accuracy of 0.55 (95\% CI, 0.46-0.64) and AUROCs ranging from 0.76 to 0.86 for the different genetic aberrations based on the internal WSIs, with slightly worse performance on the consultation WSIs. The AI models trained using the features extracted with the first and second stage of HIPT were both outperformed by the UNI-based model (see Supplementary Tables~2~and~3). For comparison, random predictions would approximately yield an accuracy of 0.25 and AUROCs of 0.50. The difference between the accuracy of the best AI model and the accuracy when randomly guessing is statistically significant (\textit{P} < 0.001). The clinical logistic regression model did not exceed random chance-level performance (see Supplementary Table~1).

Visual inspection of the attention maps for correctly and incorrectly classified cases revealed some patterns. For example, Spitz tumors predicted to harbor an \textit{NTRK} fusion regularly displayed epithelioid melanocytes in combination with pigmentation and inflammatory cells on the tiles that were assigned the largest attention weight. Cases predicted to belong to the class with other Spitz tumors frequently showed melanocytes with strong variation in cell size and pronounced nuclear atypia on these tiles. The most important tile for \textit{ALK} fusion-predicted Spitz tumors occasionally showed spindled melanocytes. It must be noted, however, that these patterns were not consistently observed across all lesions of a predicted subtype, and no clear resemblance was seen between the highest attention tiles for lesions classified to harbor a \textit{ROS1} fusion.

\begin{table}[h!]
\centering
\caption{Results for the Spitz genetic aberration prediction on the test set using the image-only AI model based on features extracted with UNI.}
\vspace{0.1cm}
\begin{tabular}{cccc}
\toprule\toprule
Metric         & Classes      &  \multicolumn{2}{c}{Performance} \\ \cmidrule(lr){3-4}
& & Internal WSIs & Consultation WSIs \\ \hline
Accuracy~(95\% CI) & \textit{ALK}, \textit{ROS1}, \textit{NTRK}, other   & 0.55~~{\scriptsize(0.46-0.64)} & 0.51~~{\scriptsize(0.41-0.60)} \\
AUROC~~(95\% CI)    & \textit{ALK} vs. rest   & 0.79~~{\scriptsize(0.67-0.89)} & 0.71~~{\scriptsize(0.56-0.84)} \\
         & \textit{ROS1} vs. rest~  & 0.76~~{\scriptsize(0.66-0.85)} & 0.77~~{\scriptsize(0.68-0.86)} \\
         & \textit{NTRK} vs. rest~~  & 0.81~~{\scriptsize(0.77-0.89)} & 0.77~~{\scriptsize(0.68-0.85)} \\
         & Other vs. rest & 0.86~~{\scriptsize(0.76-0.94)} & 0.81~~{\scriptsize(0.71-0.91)} \\ \bottomrule\bottomrule
\multicolumn{4}{l}{vs. = versus}\\
\end{tabular}
\label{tab:genetic_aberration}
\end{table}

\subsection*{Spitz Diagnostic category Prediction}
The best results for the diagnostic category prediction of Spitz tumors were achieved using features extracted with UNI, as shown in Table~\ref{tab:diagnostic_classification}. Evaluated on the internal test set WSIs, the AI model reached an accuracy of 0.51 (95\% CI, 0.40-0.60) and AUROCs of 0.62, 0.57, and 0.74 in distinguishing benign, intermediate, and malignant Spitz tumors from the rest, respectively. In contrast to the previous two prediction tasks, the difference in performance between the image feature encoders is smaller (see Supplementary Table~5~and~6) and the performance difference on the internal and consultation WSIs is less unequivocal. Random predictions would approximately yield an accuracy of 0.33 and AUROCs of 0.50. The difference between the accuracy of the best AI model and the accuracy when guessing randomly is statistically significant (\textit{P} < 0.001). Similar to the genetic aberration prediction task, the clinical logistic regression model did not exceed the performance level of random guessing (see Supplementary Table~4).

\begin{table}[h!]
\centering
\caption{Results for the Spitz diagnostic category prediction on the test set using the image-only AI model based on features extracted with UNI.}
\vspace{0.1cm}
\begin{tabular}{cccc}
\toprule\toprule
Metric         & Classes      &  \multicolumn{2}{c}{Performance} \\ \cmidrule(lr){3-4}
& & Internal WSIs & Consultation WSIs \\ \hline
Accuracy~(95\% CI) & Benign, Intermediate, Malignant          & 0.51~~{\scriptsize(0.40-0.60)} & 0.52~~{\scriptsize(0.41-0.62)} \\
AUROC~~(95\% CI)    & Benign vs. rest~~       & 0.62~~{\scriptsize(0.51-0.73)} & 0.65~~{\scriptsize(0.54-0.76)} \\
         & Intermediate vs. rest~~~~~~~~~\, & 0.57~~{\scriptsize(0.45-0.69)} & 0.62~~{\scriptsize(0.51-0.73)} \\
         & Malignant vs. rest~~~~~~    & 0.74~~{\scriptsize(0.56-0.89)} & 0.71~~{\scriptsize(0.54-0.86)} \\ \bottomrule\bottomrule
\multicolumn{4}{l}{vs. = versus}\\
\end{tabular}
\label{tab:diagnostic_classification}
\end{table}

\newpage
\subsection*{Reader Study}
The results of the reader study, comparing the performance of four pathologists experienced in dermatopathology to that of the best image-only AI models, are shown in Table~\ref{tab:reader_study}. For each of the three classification tasks, the AI model reached a higher accuracy than the four pathologists. For the first task of distinguishing Spitz tumors from conventional melanomas, the mean accuracy of the pathologists was 0.77, and the accuracy of the AI model was 0.89, with a statistically significant difference between one of the pathologists and the AI model. For the second task of predicting the genetic aberration of Spitz tumors, the mean accuracy of the pathologists and the accuracy of the AI model were 0.35 and 0.52, respectively, with no statistically significant differences in the individual comparisons. For the third task of predicting the diagnostic category of Spitz tumors, the pathologists achieved a mean accuracy of 0.36, while the AI model achieved an accuracy of 0.54, with no statistically significant differences in the individual comparisons.

\begin{table}[h!]
\centering
\caption{Results of the reader study on a randomly selected, stratified subset of the test set comparing the performance of four pathologists to that of the best image-only AI model across three tasks: distinguishing Spitz tumors from conventional melanomas, predicting the genetic aberration of Spitz tumors, and predicting the diagnostic category of Spitz tumors. The \textit{P} values were obtained using McNemar's exact test with Bonferroni-correction and pertain to the individual comparisons with the AI model for the corresponding task.}
\resizebox{\columnwidth}{!}{%
\begin{tabular}{clcccclcccclcccc}
\toprule\toprule
              &  & \multicolumn{3}{c}{Spitz tumor vs. conv. melanoma}       &  & \multicolumn{3}{c}{Genetic abberation}               &  & \multicolumn{3}{c}{Diagnostic category}                    \\ \cline{3-5} \cline{7-9} \cline{11-13} 
              &  & N   & Accuracy (95\% CI)          & \textit{P} value &  & N  & Accuracy (95\% CI)          & \textit{P} value  &  & N  & Accuracy (95\% CI)          & \textit{P} value              \\ \hline
Pathologist 1 &  & 100 & 0.81~~(0.73-0.88) & .39              &  & 43 & 0.28~~(0.16-0.40) & ~~.05   &  & 43 & 0.35~~(0.21-0.49) & >.99~~   \\
Pathologist 2 &  & 100 & 0.79~~(0.72-0.86) & .17              &  & 50 & 0.44~~(0.30-0.58) & >.99    &  & 50 & 0.40~~(0.36-0.46) & >.99~~   \\
Pathologist 3 &  & 100 & 0.71~~(0.63-0.79) & ~~.002           &  & 46 & 0.33~~(0.20-0.46) & ~~.17   &  & 46 & 0.33~~(0.24-0.41) & ~~.54~~  \\
Pathologist 4 &  & 100 & 0.76~~(0.67-0.84) & .10              &  & 39 & 0.36~~(0.23-0.49) & ~~.95   &  & 39 & 0.36~~(0.26-0.46) & >.99~~   \\ \hline
AI models     &  & 100 & 0.89~~(0.83-0.95) & -                &  & 50 & 0.52~~(0.42-0.62) & -       &  & 50 & 0.54~~(0.40-0.66) & -        \\
\bottomrule\bottomrule
\multicolumn{10}{l}{vs. = versus, conv. = conventional}\\
\end{tabular}
}
\label{tab:reader_study}
\end{table}

\subsection*{Simulation Experiment}
The results of the simulation experiment are shown in Fig.~\ref{tab:simulation_results}. Performing the Spitz IHC stains sequentially, compared to performing them in parallel, has a lower accumulated material cost, while the average turnaround time and the number of examinations are higher. Adopting AI-based recommendations, both by skipping IHC staining for Spitz tumors predicted to harbor a different genetic aberration (i.e., not \textit{ALK}, \textit{ROS1}, or \textit{NTRK}-fused) and by performing the sequential IHC stains ordered based on the predicted probability instead of the prevalence, improved the efficiency over the baseline approaches. More specifically, for the parallel IHC staining variant, the material cost accumulated over 100 cases decreased by \euro\,2,671 (3.5\%), the average turnaround time increased by 0.17 days (3.0\%), and the average number of examinations decreased by 0.18 (7.3\%). For the variant with sequential IHC staining, the material cost accumulated over 100 cases decreased by \euro\,3,996 (5.6\%), the average turnaround time decreased by 0.40 days (6.0\%), and the average number of examinations decreased by 0.76 (19.6\%). Further improvements were observed for both variants across all three metrics using the hypothetical perfect AI-based recommendations in the workflow.

\begin{table}[h!]
\centering
\caption{Results for the simulation experiment. A baseline, AI-based recommendation, and hypothetical perfect AI-based recommendation workflow were compared using parallel and sequential immunohistochemistry (IHC) assessment. The performance was measured in terms of the material cost accumulated over 100 cases, the average turnaround time per case, and the average number of examinations per case. The colors range from red to green for the maximum to minimum value per metric.}
\begin{tabular}{@{}lrcccl@{}}
\toprule\toprule
\multicolumn{2}{c}{Metric}                 & \multicolumn{1}{l}{Parallel IHC}                                                         & \multicolumn{2}{c}{Sequential IHC}                                                                                                                                                  &  \\ \cmidrule(lr){4-5}
\multicolumn{2}{c}{}                       & \multicolumn{1}{l}{}                                                                     & \multicolumn{1}{c}{Prevalence}                                                            & \multicolumn{1}{c}{Pred. prob.}                                                &  \\ \midrule
\multicolumn{2}{c}{Material cost (\euro)}            & \multicolumn{1}{l}{}                                                                     & \multicolumn{1}{l}{}                                                                     & \multicolumn{1}{l}{}                                                                     &  \\ \midrule
     & \cellcolor[HTML]{FFFFFF}Baseline     & \cellcolor[HTML]{F8696B}\begin{tabular}[c]{@{}c@{}}77,068\\ {\scriptsize(67,000-87,000)}\end{tabular} & \cellcolor[HTML]{FECA7E}\begin{tabular}[c]{@{}c@{}}71,052\\ {\scriptsize(60,200-82,200)}\end{tabular} & \cellcolor[HTML]{EFEFEF}-                                                                &  \\
     & AI-based                           & \cellcolor[HTML]{FB9474}\begin{tabular}[c]{@{}c@{}}74,397\\ {\scriptsize(65,600-83,400)}\end{tabular} & \cellcolor[HTML]{FBE983}\begin{tabular}[c]{@{}c@{}}68,742\\ {\scriptsize(58,700-78,900)}\end{tabular} & \cellcolor[HTML]{E2E282}\begin{tabular}[c]{@{}c@{}}67,056\\ {\scriptsize(56,800-77,600)}\end{tabular} &  \\
     & Perfect AI-based                   & \cellcolor[HTML]{FFE884}\begin{tabular}[c]{@{}c@{}}69,206\\ {\scriptsize(60,900-77,800)}\end{tabular} & \cellcolor[HTML]{A7D17E}\begin{tabular}[c]{@{}c@{}}63,190\\ {\scriptsize(53,700-73,000)}\end{tabular} & \cellcolor[HTML]{63BE7B}\begin{tabular}[c]{@{}c@{}}58,619\\ {\scriptsize(48,400-69,100)}\end{tabular} &  \\ \midrule
\multicolumn{2}{c}{Turnaround time (days)} & \multicolumn{1}{l}{}                                                                     & \multicolumn{1}{l}{}                                                                     & \multicolumn{1}{l}{}                                                                     &  \\ \midrule
     & Baseline                             & \cellcolor[HTML]{A2D07E}\begin{tabular}[c]{@{}c@{}}5.71\\ {\scriptsize(4.70-6.70)}\end{tabular}       & \cellcolor[HTML]{F8696B}\begin{tabular}[c]{@{}c@{}}7.11\\ {\scriptsize(6.02-8.22)}\end{tabular}       & \cellcolor[HTML]{EFEFEF}-                                                                &  \\
     & AI-based                           & \cellcolor[HTML]{CBDC81}\begin{tabular}[c]{@{}c@{}}5.88\\ {\scriptsize(4.93-6.84)}\end{tabular}       & \cellcolor[HTML]{FA8871}\begin{tabular}[c]{@{}c@{}}6.87\\ {\scriptsize(5.87-7.89)}\end{tabular}       & \cellcolor[HTML]{FB9D75}\begin{tabular}[c]{@{}c@{}}6.71\\ {\scriptsize(5.68-7.76)}\end{tabular}       &  \\
     & Perfect AI-based                   & \cellcolor[HTML]{63BE7B}\begin{tabular}[c]{@{}c@{}}5.44\\ {\scriptsize(4.53-6.40)}\end{tabular}       & \cellcolor[HTML]{FECF7F}\begin{tabular}[c]{@{}c@{}}6.32\\ {\scriptsize(5.37-7.30)}\end{tabular}       & \cellcolor[HTML]{C6DA80}\begin{tabular}[c]{@{}c@{}}5.86\\ {\scriptsize(4.84-6.91)}\end{tabular}       &  \\ \midrule
\multicolumn{2}{c}{Number of examinations} & \multicolumn{1}{l}{}                                                                     & \multicolumn{1}{l}{}                                                                     & \multicolumn{1}{l}{}                                                                     &  \\ \midrule
     & Baseline                             & \cellcolor[HTML]{A1D07E}\begin{tabular}[c]{@{}c@{}}2.47\\ {\scriptsize(2.37-2.57)}\end{tabular}       & \cellcolor[HTML]{F8696B}\begin{tabular}[c]{@{}c@{}}3.87\\ {\scriptsize(3.63-4.11)}\end{tabular}       & \cellcolor[HTML]{EFEFEF}-                                                                &  \\
     & AI-based                           & \cellcolor[HTML]{76C37C}\begin{tabular}[c]{@{}c@{}}2.29\\ {\scriptsize(2.20-2.38)}\end{tabular}       & \cellcolor[HTML]{FDB57A}\begin{tabular}[c]{@{}c@{}}3.28\\ {\scriptsize(3.03-3.53)}\end{tabular}       & \cellcolor[HTML]{FECB7E}\begin{tabular}[c]{@{}c@{}}3.11\\ {\scriptsize(2.86-3.37)}\end{tabular}       &  \\
     & Perfect AI-based                   & \cellcolor[HTML]{63BE7B}\begin{tabular}[c]{@{}c@{}}2.21\\ {\scriptsize(2.13-2.29)}\end{tabular}       & \cellcolor[HTML]{FECF7F}\begin{tabular}[c]{@{}c@{}}3.08\\ {\scriptsize(2.85-3.32)}\end{tabular}       & \cellcolor[HTML]{C8DB80}\begin{tabular}[c]{@{}c@{}}2.63\\ {\scriptsize(2.39-2.87)}\end{tabular}       &  \\ \bottomrule\bottomrule
     \multicolumn{6}{l}{Pred. prob. = Predicted probability}\\
\end{tabular}
\label{tab:simulation_results}
\end{table}

\section*{Discussion and Conclusion}
In this study, we investigated the extent to which an AI model can accurately distinguish Spitz tumors from conventional melanomas and predict the underlying genetic aberration and diagnostic category of Spitz tumors. We conducted a reader study to compare the predictive performance of AI models with that of four pathologists on these tasks. Additionally, to better understand how AI-based recommendations for ancillary diagnostic testing could affect the workflow of the pathology department, we performed a simulation experiment.

The best AI model correctly distinguished most Spitz tumors from conventional melanomas, as evidenced by an AUROC of 0.95 and an accuracy of 0.86 on the test set. The classification performance varied between feature extraction models, with the second stage of HIPT performing better than the first stage, while both were outperformed by UNI. These findings align with previously reported results for classification tasks in other pathology domains~\cite{chen2022scaling,campanella2025clinical,xu2024whole}. Our results showed that a logistic regression model based solely on age, sex, and anatomical location performed reasonably well; however, using these clinical features in combination with the best image-based prediction model did not improve performance. This is noteworthy, as pathologists typically do heavily rely on clinical information when diagnosing Spitzoid lesions. Slightly lower performance was observed in the evaluation based on the consultation WSIs, which can likely be attributed to the variation in tissue appearance due to differences in preparation and staining protocols between centers~\cite{van2021deep}. Moreover, the presence of nevus cells is a relevant histological feature for diagnosis, as these cells are regularly seen together with conventional melanomas (i.e., in the form of a pre-existing nevus), while a nevus next to a Spitz tumor is very uncommon~\cite{meijs2025stranger}. The attention visualization suggests that the AI model has also learned to recognize this characteristic (Fig~\ref{fig:attention_maps}A, center column).

For predicting the genetic aberration, the best AI model reached a classification performance significantly above random chance-level, reaching an accuracy of 0.55, where random predictions would yield 0.25. Visual inspection of the tiles with the highest attention weights revealed some patterns consistent with characteristics described in case studies of Spitz tumors with specific genetic aberrations~\cite{gerami2021clinical,yeh2015clinical,yeh2019filigree,sharma2024clinical,patel2024clinical,de2025spitz}, although interpretation remained challenging. Incorporating positional embeddings can potentially further improve classification performance by enabling the AI model to also capture the lesion morphology at lower magnification as well.

The diagnostic category prediction was the most challenging task, as the best AI model achieved an accuracy of 0.51, compared to 0.33 for random guessing. To reach a diagnostic category for Spitz tumors in clinical practice, pathologists need to integrate histological, immunohistochemical, and genetic features to arrive at a diagnosis, without strict criteria for which feature combinations constitute a Spitz nevus, melanocytoma, or melanoma~\cite{who2023}. Despite the improvement in agreement between experts with the availability of genetic information, disagreement remained in a considerable fraction of cases~\cite{benton2021impact}, illustrating the difficulty of diagnosing Spitzoid lesions. This diagnostic variability may have affected the model development and evaluation. Nevertheless, the limited predictive performance is likely primarily due to the absence of histological characteristics that correlate with the genomic background.

The reader study showed that the AI model for each of the three Spitz classification tasks reached a higher accuracy than the four pathologists with experience in dermatopathology, although the difference in accuracy was not statistically significant for most individual comparisons. It is important to note that pathologists in clinical practice typically rely on IHC stains and molecular diagnostics to differentiate Spitz tumors from conventional melanomas, and to determine the underlying genetic aberration and diagnostic category. It should therefore be expected that most pathologists are not used to performing these tasks without additional diagnostic information being available. Other factors which could have affected the pathologists' assessment include: (1) the cases were randomly selected with stratification to obtain mostly balanced classes for each of the three tasks, which ensured adequate representation of rare classes for evaluation purposes, but also resulted in class distributions that deviated from the real-world prevalences (e.g., Spitz melanomas are much more rare than Spitz nevi); (2) the pathologists performed all three tasks at once, while separate AI models were trained for the respective tasks; and (3) the WSI appearance and viewing application likely differed from the routine setup of the pathologists.

Through a simulation experiment, we studied how implementing AI models for predicting genetic aberrations might impact the workflow of the pathology department. While the accuracy is currently not high enough to serve as a replacement for IHC staining or molecular analyses, we demonstrated that AI-based recommendations on the selection of ancillary diagnostic tests can potentially improve workflow efficiency by reducing the total material cost, the turnaround times, and the number of examinations. Although the genetic background of Spitz tumors can also be predicted by pathologists and does not necessarily require an AI model, this task is challenging, as seen in the reader study, and is not routinely performed in clinical practice at the moment. The AI model could, therefore, serve as a tool for pathologists to reach the correct diagnosis faster while reducing costs. The scope of the simulation experiment was limited to Spitz tumors, which does not completely reflect clinical practice where melanocytic lesions can also be other subtypes, but does show how efficiency gains could be achieved while keeping the simulation complexity manageable. Further improvement of the predictive accuracy would yield larger gains in the direction of the hypothetical perfect AI model. Extending this approach to other relevant IHC stains for melanocytic lesions (e.g., BRAF, BAP1, $\beta$-catenin) of even other tumor types could also increase the benefits~\cite{cifci2022artificial}. In addition, simulation can also be useful for investigating the level of accuracy required in terms of expected savings to justify the costs of AI model implementation.

Despite this being the largest study into AI-based classification of Spitz tumors, the dataset size remains still comparatively small, and improvements in model performance may be possible after training on more data. Additionally, only Spitz tumor or conventional melanoma cases confirmed by a positive IHC stain for a Spitz marker and/or molecular analysis were included in the study cohort. This inclusion criterion has likely introduced some form of selection bias, as conventional melanomas are not always genetically characterized in routine practice, nor do all harbor a \textit{BRAF} or \textit{NRAS} mutation. Improvements in molecular diagnostic equipment have also enabled the identification of more Spitz subtypes over time. In combination with the specialized caseload as consultation center, this could have resulted in prevalences that differ from those in the general population.

In conclusion, the AI model achieved a strong predictive performance in distinguishing Spitz tumors from conventional melanomas. On the more challenging tasks of predicting the genetic aberration and the diagnostic category of Spitz tumors, the AI models performed better than random chance. The potential benefits of implementing AI-based recommendations for ancillary diagnostic testing were demonstrated using a simulation experiment.

\section*{Author Contribution}
R.L., M.V., and W.B. conceptualized the study. 
R.L., M.R., C.E., A.N., N.S., G.B., A.J., and W.B. participated in data curation and verification. 
R.L. and M.V. designed the methodology.
R.L. developed the AI models and performed the model evaluation.
A.M, D.H., L.W., S.R. participated in the reader study.
R.L., M.R., N.S. performed the reader study evaluation.
R.L., M.R., M.V., W.B. analyzed and interpreted the results.
R.L. wrote the original draft.
M.V. and W.B. supervised the project and participated in funding acquisition.
All authors had full access to all the data in the study.
All authors read, edited, and approved the final manuscript.
All authors accept the final responsibility to submit for publication and take responsibility for the contents of the manuscript.

\section*{Data Availability}
All relevant data supporting the findings of this study are available within the paper and its Supplementary Information. Raw data that support the findings of this study are not openly available because of patient privacy reasons, but can be made available upon reasonable request. Requests for access can be directed to the corresponding author.

\section*{Funding}
This research was financially supported by the Hanarth Fonds. The funder played no role in study design, data collection, analysis and interpretation of data, or the writing of this manuscript.

\section*{Declaration of Competing Interest}
The authors declare no competing interests. 

\section*{Ethics Approval and Consent to Participate}
The study does not fall within the scope of the Dutch Medical Research Involving Human Subjects Act (WMO) and therefore does not require approval from an accredited medical ethics committee in the Netherlands. Nevertheless, an independent quality assessment (25U-0162) was conducted at the UMC Utrecht to ensure compliance with relevant laws and regulations, including those related to the informed consent procedure, data management, privacy, and legal considerations. The need for informed consent was waived due to the cohort size and retrospective nature of the study. 

\printbibliography

\renewcommand{\tablename}{Supplementary Table}
\renewcommand{\figurename}{Supplementary Figure}
\setcounter{table}{0}
\setcounter{figure}{0}

\newpage
\section*{Supplementary Material}

\begin{table}[H]
\caption{Results for Spitz genetic aberration prediction using the logistic regression model with clinical features only.}
\centering
\begin{tabular}{ccc}
\toprule\toprule
Metric         & Classes      & Performance \\ \hline
Accuracy~(95\% CI) & \textit{ALK}, \textit{ROS1}, \textit{NTRK}, other   & 0.22~~{\scriptsize(0.14-0.30)} \\
AUROC~~(95\% CI)    & \textit{ALK} vs. rest                              & 0.61~~{\scriptsize(0.44-0.77)} \\
         & \textit{ROS1} vs. rest~                                       & 0.50~~{\scriptsize(0.38-0.62)} \\
         & \textit{NTRK} vs. rest~~                                      & 0.39~~{\scriptsize(0.28-0.51)} \\
         & Other vs. rest                                                & 0.56~~{\scriptsize(0.41-0.70)} \\ \bottomrule\bottomrule
\multicolumn{3}{l}{vs. = versus}\\
\end{tabular}
\label{tab:genetic_aberration_clin}
\end{table}

\begin{table}[H]
\caption{Results for Spitz genetic aberration prediction using the image-only AI model based on features extracted with the first stage of HIPT.}
\centering
\begin{tabular}{cccc}
\toprule\toprule
Metric         & Classes      &  \multicolumn{2}{c}{Performance} \\ 
& & Internal WSIs & Consultation WSIs \\ \hline
Accuracy~(95\% CI) & \textit{ALK}, \textit{ROS1}, \textit{NTRK}, other    & 0.30~~{\scriptsize(0.22-0.38)} & 0.29~~{\scriptsize(0.23-0.35)} \\
AUROC~~(95\% CI)    & \textit{ALK} vs. rest                               & 0.66~~{\scriptsize(0.50-0.81)} & 0.54~~{\scriptsize(0.38-0.68)} \\
         & \textit{ROS1} vs. rest~                                        & 0.50~~{\scriptsize(0.37-0.62)} & 0.57~~{\scriptsize(0.44-0.69)} \\
         & \textit{NTRK} vs. rest~~                                       & 0.64~~{\scriptsize(0.52-0.76)} & 0.62~~{\scriptsize(0.51-0.73)} \\
         & Other vs. rest                                                 & 0.60~~{\scriptsize(0.46-0.72)} & 0.57~~{\scriptsize(0.44-0.70)} \\ \bottomrule\bottomrule
\multicolumn{4}{l}{vs. = versus}\\
\end{tabular}
\label{tab:genetic_aberration_hipt1}
\end{table}

\begin{table}[H]
\caption{Results for Spitz genetic aberration prediction using the image-only AI model based on features extracted with the second stage of HIPT.}
\centering
\begin{tabular}{cccc}
\toprule\toprule
Metric         & Classes      &  \multicolumn{2}{c}{Performance} \\ 
& & Internal WSIs & Consultation WSIs \\ \hline
Accuracy~(95\% CI) & \textit{ALK}, \textit{ROS1}, \textit{NTRK}, other   & 0.39~~{\scriptsize(0.30-0.49)} & 0.36~~{\scriptsize(0.28-0.46)} \\
AUROC~~(95\% CI)    & \textit{ALK} vs. rest                              & 0.67~~{\scriptsize(0.53-0.80)} & 0.56~~{\scriptsize(0.40-0.72)} \\
         & \textit{ROS1} vs. rest~                                       & 0.63~~{\scriptsize(0.51-0.74)} & 0.67~~{\scriptsize(0.54-0.79)} \\
         & \textit{NTRK} vs. rest~~                                      & 0.68~~{\scriptsize(0.56-0.79)} & 0.67~~{\scriptsize(0.55-0.78)} \\
         & Other vs. rest                                                & 0.70~~{\scriptsize(0.58-0.82)} & 0.66~~{\scriptsize(0.52-0.78)} \\ \bottomrule\bottomrule
\multicolumn{4}{l}{vs. = versus}\\
\end{tabular}
\label{tab:genetic_aberration_hipt2}
\end{table}

\begin{table}[H]
\caption{Results for Spitz diagnostic classification prediction using the logistic regression model with clinical features only.}
\centering
\begin{tabular}{ccc}
\toprule\toprule
Metric         & Classes      & Performance \\ \hline
Accuracy~(95\% CI) & Benign, Intermediate, Malignant   & 0.35~~{\scriptsize(0.26-0.45)} \\
AUROC~~(95\% CI)    & Benign vs. rest~~                & 0.44~~{\scriptsize(0.32-0.56)} \\
         & Intermediate vs. rest~~~~~~~~~\,            & 0.63~~{\scriptsize(0.52-0.73)} \\
         & Malignant vs. rest~~~~~~                    & 0.52~~{\scriptsize(0.35-0.68)} \\ \bottomrule\bottomrule
\multicolumn{3}{l}{vs. = versus}\\
\end{tabular}
\label{tab:diagnostic_classification_clin}
\end{table}

\begin{table}[H]
\centering
\caption{Results for Spitz diagnostic classification prediction using the image-only AI model based on features extracted with the first stage of HIPT.}
\begin{tabular}{cccc}
\toprule\toprule
Metric         & Classes      &  \multicolumn{2}{c}{Performance} \\ 
& & Internal WSIs & Consultation WSIs \\ \hline
Accuracy~(95\% CI) & Benign, Intermediate, Malignant          & 0.48~~{\scriptsize(0.37-0.58)} & 0.41~~{\scriptsize(0.32-0.52)} \\
AUROC~~(95\% CI)    & Benign vs. rest~~       & 0.63~~{\scriptsize(0.52-0.74)} & 0.64~~{\scriptsize(0.52-0.74)} \\
         & Intermediate vs. rest~~~~~~~~~\, & 0.52~~{\scriptsize(0.40-0.63)} & 0.53~~{\scriptsize(0.41-0.65)} \\
         & Malignant vs. rest~~~~~~    & 0.72~~{\scriptsize(0.59-0.83)} & 0.73~~{\scriptsize(0.59-0.84)} \\ \bottomrule\bottomrule
\multicolumn{4}{l}{vs. = versus}\\
\end{tabular}
\label{tab:diagnostic_classification_hipt1}
\end{table}

\begin{table}[H]
\centering
\caption{Results for Spitz diagnostic classification prediction using the image-only AI model based on features extracted with the second stage of HIPT.}
\begin{tabular}{cccc}
\toprule\toprule
Metric         & Classes      &  \multicolumn{2}{c}{Performance} \\ 
& & Internal WSIs & Consultation WSIs \\ \hline
Accuracy~(95\% CI) & Benign, Intermediate, Malignant          & 0.50~~{\scriptsize(0.39-0.60)} & 0.46~~{\scriptsize(0.36-0.55)} \\
AUROC~~(95\% CI)    & Benign vs. rest~~       & 0.66~~{\scriptsize(0.55-0.76)} & 0.68~~{\scriptsize(0.57-0.78)} \\
         & Intermediate vs. rest~~~~~~~~~\, & 0.56~~{\scriptsize(0.44-0.67)} & 0.62~~{\scriptsize(0.51-0.73)} \\
         & Malignant vs. rest~~~~~~    & 0.71~~{\scriptsize(0.55-0.85)} & 0.79~~{\scriptsize(0.68-0.89)} \\ \bottomrule\bottomrule
\multicolumn{4}{l}{vs. = versus}\\
\end{tabular}
\label{tab:diagnostic_classification_hipt2}
\end{table}

\end{document}